\documentclass[aps,pre,twocolumn,showpacs]{revtex4}

\newcommand{\eps}{\varepsilon}

\usepackage{epsfig}
\usepackage{graphicx}
\usepackage{amsmath}
\usepackage{amssymb}
\usepackage{dcolumn}

\begin{document}

\draft

\title {Stable and unstable vector dark solitons of coupled nonlinear Schr\"{o}dinger equations. Application to two-component Bose-Einstein condensates}

\author{V. A. Brazhnyi$^1$}
\email{brazhnyi@cii.fc.ul.pt}
\author{V. V. Konotop$^{1,2}$}
\email{konotop@cii.fc.ul.pt}

\address{
$^1$ Centro de F\'{\i}sica Te\'orica e Computacional,
Universidade de Lisboa,
    Complexo Interdisciplinar, Av. Prof. Gama Pinto 2, Lisboa 1649-003,
         Portugal 
\\
$^2$ Departamento de F\'{\i}sica,
Universidade de Lisboa, Campo Grande, Ed. C8, Piso 6, Lisboa
1749-016, Portugal
          }

\date{\today}


\begin{abstract}
Dynamics of vector dark solitons in two-component Bose-Einstein condensates is studied within the framework of the coupled one-dimensional nonlinear Schr\"odinger (NLS) equations. We consider the small amplitude limit in which the coupled NLS equations are reduced to the coupled Korteweg-de Vries (KdV) equations.
For a specific choice of the parameters the obtained coupled KdV equations are exactly integrable. We find that there exist two branches of (slow and fast) dark solitons corresponding to the two branches of the sound waves. Slow solitons, corresponding to the lower branch of the acoustic wave appear to be unstable and transform during the evolution into the stable fast solitons (corresponding to the upper branch of the dispersion law). Vector dark solitons of arbitrary depths are studied numerically. It is shown that effectively different parabolic traps, to which the two components are subjected, cause instability of the solitons leading to splitting of their components and subsequent decay. Simple phenomenological theory, describing oscillations of vector dark solitons in a magnetic trap is proposed.
\end{abstract}

\maketitle

\section{Introduction}

It is well established that in a one-component Bose-Einstein condensate (BEC) with a positive scattering length, which has cigar-shaped geometry, one can generate dark solitons~\cite{dark}. Experimental generation of two-component  BEC's of different hyperfine states of rubidium atoms in a magnetic trap~\cite{exp1} and of sodium atoms in an optical trap \cite{exp2} stimulated theoretical studies devoted to the meanfield dynamics of multicomponent condensates. As in the one-component case special attention was devoted to existence of solitary waves in such systems. When a condensate is cigar-shaped and has relatively low density, i.e. when the healing lengths of the components are much larger than the transverse dimension of the condensate and much less than its longitudinal dimension,  the transverse atomic distribution is well approximated by the Gaussian ground state and the system of coupled Gross-Pitaevskii (GP) equations, describing the mixture (see e.g. \cite{PetSmith}),  can be reduced to the coupled one-dimensional (1D) nonlinear Schr\"{o}dinger (NLS) equations (see e.g.~\cite{Kostov}). The respective models were a subject of recent theoretical studies. In particular, coupled  large-amplitude dark-bright solitons have been reported in~\cite{BA}; bound  dark solitons have been numerically studied in \cite{OS}, where it has been found that creation of slowly moving objects is possible; a diversity of other bound states has been generated numerically in \cite{KNF}. 

The present paper aims further analytical and numerical study of dark solitons in  two-component BEC's.  Main distinctions of the situation considered here compared to the previous research are as follows. (i) We consider {\em vector} dark solitons, i.e. states where two components move with equal or approximately equal velocities. (ii) We do not impose the condition of equality of the nonlinear coefficients, as a necessary condition, allowing one to reduce the problem to the exactly integrable one -- to the so-called Manakov problem, for which vector dark solitons are known \cite{RL}. (iii) In the small amplitude limit we provide  analytical description of the phenomenon reducing a system of coupled NLS equations to a system of coupled Korteweg-de Vries (KdV) equations, what allows us to predict existence of two types of vector dark solitons, moving with different velocities. (iv) Finally, we study in detail the effect of the magnetic trap on the dark soliton dynamics. We show that, due to difference of its effect on different component, a magnetic trap leads to splitting of the components and subsequent destruction of vector dark solitons.

\section{Statement of the problem and physical parameters}

Evolution of a two-component BEC composed of different hyperfine states is described  by the coupled GP equations ($j=1,2$)~\cite{PetSmith}
\begin{eqnarray}
\label{GP}
i\hbar\frac{\partial \Psi_j}{\partial t}=\left[-\frac{\hbar^2}{2m}\nabla^2+V_j({\bf r})+\frac{4\pi\hbar^2}{m}\sum\limits_{l=1,2}a_{jl}|\Psi_l|^2\right]\Psi_j 
\end{eqnarray}
where $
\displaystyle{V_j({\bf r})=\frac{m}{2}\omega_j^2(\lambda^2x^2+r_\bot^2)}
$, $\Psi_j$ are the macroscopic wave functions of the states, $a_{ij}$ are  scattering lengths of the respective interactions -- it will be assumed that they are positive, $\omega_j$ are transverse linear oscillator frequencies of the components, and $\lambda$ is the aspect ratio of the condensate. Respectively, ${\cal N}_j=\int|\Psi_j|^2d{\bf r}$ is the number of atoms of $j$-th component and ${\cal N}={\cal N}_1+{\cal N}_2$ is the total number of atoms. 

In the case of an elongated trap, when $\lambda$ is small enough, and when densities of the both components are low enough, one can employ the multiple-scale expansion method in order to reduce the original 3D system (\ref{GP}) to the homogeneous coupled 1D GP equations (for details of derivation see e.g.~\cite{Kostov})
\begin{eqnarray}
\label{GP1}
\begin{array}{l}
\displaystyle{
i\partial_T \Phi_1 =- \partial_X^2 \Phi_1 + \chi_{1}|\Phi_1|^2\Phi_1 + \chi|\Phi_2|^2\Phi_1,
}
\\ 
\displaystyle{
i \partial_T \Phi_2 =-\partial_X^2 \Phi_2 + \chi|\Phi_1|^2\Phi_2 + \chi_{2}|\Phi_2|^2\Phi_2.
}
\end{array}
\end{eqnarray}
Here  
\begin{eqnarray}
\label{chi}
\begin{array}{l}
\displaystyle{
\chi_{1}= \frac{1}{(2\pi)^{3/2}}, 
\qquad
\chi_{2}= \left(\frac{\omega}{2\pi}\right)^{3/2}\frac{a_{22}}{|a_{11}|},
}
\\
\displaystyle{
\chi = \left(\frac{\omega}{\pi(1+\omega)}\right)^{3/2}\frac{a_{12}}{|a_{11}|},
}
\end{array}
\end{eqnarray}
$\Phi_j$,  $T$ and $X$ are the dimensionless wave function envelops, slow time and slow coordinate, respectively, $\omega=\omega_2/\omega_1$, 
$g_{ij}=\frac{4\pi {\cal N} a_{ij}}{a_1}$, and $a_j=\sqrt{\frac{\hbar}{m\omega_j}}$. The small parameter of the problem is defined as 
\begin{eqnarray}
\label{epsilon}
\delta=\sqrt{\frac{8\pi {\cal N}_1a_{11}\lambda^{1/2}}{a_1}}. 
\end{eqnarray}

Besides asymmetry of the trap and weakness of the two-body interactions, expressed by the smallness of $\lambda$ and $\delta$, respectively, our model assumes equality of the aspect ratios of the components ($\lambda$ does not depend on $j$). Meantime we emphasize that the linear oscillator frequencies $\omega_j$ may be very different (say, in the experiment reported in Ref.~\cite{MMFMI}, the relation between the frequencies was $\omega=\sqrt{2}$). As a result, even for initially equal scattering lengths, the effective nonlinearities $\chi_j$, defined in Eq.~(\ref{chi}),  become different because of different transverse distributions of the components.  

To estimate a typical value of the parameter $\delta$ we consider a binary condensate of two hyperfine states of rubidium atoms in a trap with the mean value of the transverse oscillator frequency $2\pi\times 200$~Hz and the aspect ratio $\lambda=10^{-4}$ (this corresponds to the 1~$\mu$m and 100~$\mu$m of the transverse and longitudinal linear oscillator lengths). 
Taking $a_{11}\approx 1$~nm (here we take into account that the scattering length can be varied by using Feshbach resonance) and assuming the mean atomic density to be $n\approx 10^{12}\,$cm$^{-3}$ we obtain $\delta\approx 0.05$. We also point out here that the respective healing length $\xi$ (for numerical estimates we assume that the healing lengths of the both components are approximately equal) is estimated to be of order of $4\,\mu$m.

\section{Small amplitude dark solitons}
 
\subsection{Rescaled system of equations}  
 
An initial value problem for system  (\ref{GP1}) does not allow solution in a general case, except in the special limit, which is known as the Manakov system and which is discussed below [see (\ref{manakov}) the respective discussion]. Some information about possible solutions, is however available in the small amplitude limit, where the coupled NLS equations are reduced to the coupled KdV equations.   

In this approximation a dark soliton evolves against a background what means that (\ref{GP1}) is considered subject to the boundary conditions
\begin{eqnarray}
\label{bound1}
\lim_{|x|\to\infty}|\Phi_j|^2=\rho_j^2
\end{eqnarray}
where $\rho_j^2$ are properly normalized densities of the components. 
Taking this into account as well as a large number of free parameters of the problem,   is convenient to scale out the boundary conditions for the sake of performing the small-amplitude reduction mentioned above. To this end we introduce the total dimensionless density $\rho^2=\rho_1^2+\rho_2^2$, and rescale the variables as follows: $\psi_j=\frac{1}{\rho_j}\Phi_j$, $t=\chi \rho^2 T$, and $x=\sqrt{\chi}\rho X$. Then Eq.~(\ref{GP1}) is rewritten in the form
\begin{eqnarray}
	\label{NLS-dimless}
	\begin{array}{l}
	i\partial_t\psi_1=-\partial_{x}^2\psi_{1}+(U_1|\psi_1|^2+\cos^2\alpha|\psi_2|^2)\psi_1\, ,
	\\
	i\partial_t   
	\psi_{2}=-\partial_{x}^2\psi_{2}+(\sin^2\alpha|\psi_1|^2+U_2|\psi_2|^2)\psi_2
	\end{array}
\end{eqnarray}
where $U_j=\frac{\chi_j\rho_j^2}{\chi \rho^2}$, $\cos \alpha=\frac{\rho_2}{\rho}$, $\sin \alpha=\frac{\rho_1}{\rho}$, and thus the parameter $\alpha$ determines the relation between the unperturbed densities of the components: 
$\tan \alpha=\frac{\rho_1}{\rho_2}$. 
The boundary conditions now acquire the desirable form
\begin{eqnarray}
\label{boundary}
\lim_{|x|\to\infty}|\psi_j|^2=1.  
\end{eqnarray}

\subsection{Sound propagation}

Small amplitude dark solitons, which from the physical point of view represent 
packets of acoustic waves, for which weak nonlinearity and weak dispersion are balanced, propagate against a background with a speed close to the group velocity of the sound (see e.g. \cite{ZK,BKP}, as well as consideration below).  The background in our case is computed from (\ref{NLS-dimless}) and (\ref{boundary}) to be $\psi_j=\exp(-i{\cal E}_jt)$ where
\begin{eqnarray}
\label{CP}
{\cal E}_1=U_1+\cos^2\alpha, \qquad {\cal E}_2=U_2+\sin^2\alpha\,.
\end{eqnarray}
Designating frequency and wave vector of a sound wave as $\Omega$ and $K$, respectively, i.e. considering a solution of (\ref{NLS-dimless}) in a form $\psi_j=1+b_j\exp\left[i(Kx-\Omega t)\right]+c_j\exp\left[-i(Kx-\Omega t)\right]$, where $b_j$ and $c_j$ are small constants, $|b_j|\,, |c_j|\ll 1$, one finds the two branches (upper with sign ``$+$'' and lower with sign ``$-$'') of the spectrum of the acoustic waves
\begin{eqnarray}
\label{omega}
	\Omega_{\pm}=K\sqrt{K^2+U_1+U_2\pm\sqrt{(U_2-U_1)^2+\sin^2(2\alpha)}}.
\end{eqnarray}
Since dark solitons will be constructed against a static background we are interested in the limit of long wavelengths, where the group velocities are given by:
\begin{eqnarray}
	\label{group}
	v_\pm &=& \lim_{K\to 0}\frac{d\Omega_\pm}{dK}
	\nonumber \\
	&=& \sqrt{U_2+U_1\pm\sqrt{(U_2-U_1)^2+\sin^2(2\alpha)}}.
\end{eqnarray}
It follows from (\ref{omega}), that the lower branch is stable subject to the condition 
\begin{eqnarray}
\label{cond_stab}
	\Delta = U_1U_2- \sin^2\alpha\cos^2\alpha\geq 0
\end{eqnarray}
and thus, the consideration below will be restricted only to this case. We notice that  this constrain corresponds to the condition of the thermodynamic stability of the condensate with the only difference, that it is written for the effective nonlinearities (rather than in terms of the coefficients $g_{ij}$, see e.g.~\cite{PetSmith}).

An interesting feature, relevant to the next consideration, is that the group velocity of the long-length excitations of the lowest branch, i.e. $v_-$, becomes zero at
\begin{eqnarray}
	\label{manakov}
	\sin^2(2\alpha)=4U_1U_2.
\end{eqnarray}
 Then the matrix of effective nonlinearities $\Delta$ is degenerated. This situation corresponds to the Manakov system, which is an integrable limit of the coupled NLS equations. 

\subsection{A comment on the small parameter} 
 
Let us now turn to the analysis of small amplitude dark solitons. To this end we employ the idea due to Ref.~\cite{ZK} about possibility  of the multiple-scale reductions between NLS and KdV equation, noticing that the KdV equation can be obtained as a small amplitude limit of the NLS like equation also in nonintegrable limit and with arbitrary (intensity dependent) nonlinearity~\cite{BKP}. 

Before going into details of calculations we make the following observation. Derivation of the dynamical equation for small amplitude waves is based on introduction of the small parameter of the problem, which we will be designated as $\eps$ [see (\ref{expan1}), (\ref{expan2}) below]. Then the resulting evolution equation (see  (\ref{kdv}) below) appears in the order $\eps^5$. If we now recall that the 1D reduction of the coupled GP equations is mathematically justified by smallness of $\delta$, introduced in (\ref{epsilon}), and that the respective 1D NLS equations  (\ref{GP1}) appear in the $\delta^3$ order, while terms of order of $\delta^4$ are neglected, we conclude that strictly speaking the resulting KdV limit is not applicable for the description of low-dimensional BEC's [it would be valid only if one could provide the inequality $\delta \ll \eps^5\ll 1$, what is not feasible for real experimental situation, where typical values of $\delta$ are of order of $0.1\div 0.01$, as it has been mentioned above]. Thus the present section although being interested from the viewpoint of the small-amplitude limit of the coupled NLS equations, cannot be directly interpreted as a theory of small amplitude dark BEC solitons (the issue which remains to be an open problem). The significance of the results obtained below for the mean-field theory of the BEC is in the indication on new types of the solutions which cannot be directly obtained from the system (\ref{GP1}) (or (\ref{NLS-dimless})).

There exists one more limitation for the practical use of the small amplitude limit. 
It is related to the fact that small amplitude solitons are wide and their width may be comparable with the longitudinal extension of the condensate. 
Indeed, the characteristic size of a (non-small-amplitude) dark soliton is of order of a healing length $\xi_j$ (as above healing lengths of the both components $\xi_{1,2}$ are considered of the same order).  
This means that a small amplitude soliton has a width of order of $\xi_j/\varepsilon\gg\xi_j$.  
The small amplitude expansion fails at the boundaries of the atomic cloud, as the density approaches zero in those domains. 
Thus, for the validity of the theory one must require the longitudinal size of the condensate (i.e. $a_j/\sqrt{\lambda}$) to be much bigger than the width of the soliton $a_j/\sqrt{\lambda}\gg \xi_j/\varepsilon$ or in other words one must impose a condition $\varepsilon\gg\sqrt{\lambda} \xi_j/a_j$. 
The obtained constrain is , however, not as strong as the previous one. In particular, the estimates provided the end of the Sec. II give now  $\varepsilon\gg 0.04 $.

\subsection{Coupled KdV equations}

We look for a solution of system (\ref{NLS-dimless}) in a form of a small amplitude excitation of the background $\exp(-i{\cal E}_jt)$, which moves with a velocity close to one of the speeds $v_\pm$ given by (\ref{group}). The respective analytical ansatz reads
\begin{eqnarray}
\label{sol_expan}
\psi_j=Q_j(\zeta,\tau)\exp\left(-i{\cal E}_jt+i\varphi_j(\zeta,\tau)\right)
\end{eqnarray}
where the amplitude, $Q_j(\zeta,\tau)$, and the phase, $\varphi_j(\zeta,\tau)$, are represented in forms of the expansions  
\begin{eqnarray}
	\label{expan1}
	&&Q_j(\zeta,\tau)=1+\eps^2 q_{j}(\zeta,\tau)+\eps^4 q_{j1}(\zeta,\tau)+\cdots \, ,
	\\
		\label{expan2}
	&& \varphi_j(\zeta,\tau)=\eps \phi_{j}(\zeta,\tau)+\eps^3 \phi_{j1}(\zeta,\tau)+\cdots 
\end{eqnarray}
and the new slow variables are given by
\begin{eqnarray*}
&&\zeta=\eps (x-vt),
	\\
	 && \tau= \frac{\eps^3}{(U_1+U_2)[2(U_1+U_2)v^2-4\Delta]}t. 
\end{eqnarray*}
Hereafter $v$ is either $v_+$ or $v_-$, depending on the branch under consideration. One verifies, that subject to condition (\ref{cond_stab}) the denominator in the definition of the slow time $\tau$, which is introduced for the sake of convenience, is always positive. 

Our aim now is to derive evolution equations for $q_j(\zeta,\tau)$ and $\phi_j(\zeta,\tau)$, which will describe evolution of the small amplitude dark solitons. Respectively, we impose the boundary conditions
\begin{eqnarray}
\label{boundary1}
	\lim_{|\zeta|\to\infty}q_j(\zeta,\tau)=0,\quad \lim_{\zeta\to\pm\infty}\phi_j(\zeta,\tau)=\phi_{j\pm},
\end{eqnarray}
with $\phi_{j\pm}$ being constants, and consider equations of different orders of $\varepsilon$.  

While equations of the zero and the first orders are satisfied identically, it follows from the equations of the 2-nd and 3-rd orders  with respect to $\eps$ (see Appendix A for the details), that the amplitudes and the phases of the excitations of the two components are linked by the relations ($j=1,2$)
\begin{eqnarray}
	\label{amp-phase}
	q_{j}=\frac 1v\partial_\zeta\phi_{j}.
\end{eqnarray}
The obtained formula reveals essential difference between integrable and nonintegrable versions of the coupled NLS equations. In the former case a zero value of the group velocity of the lowest branch does not allow existence of small amplitude solitary pulses. In other words in the integrable case there exists only one branch of dark solitons. 

The condition of compatibility of the equations of 4-th and 5-th orders in the small parameter $\varepsilon$ results in the coupled KdV equations (see Appendix A):
\begin{eqnarray}
	\label{kdv}
	\partial_\tau q_j+
	\partial_\zeta\gamma_j^{kl}q_kq_l+\partial_{\zeta}^3\beta_j^kq_k=0 
\end{eqnarray}
 where $j,k,l=1,2$
\begin{eqnarray*}
	&&\gamma_1^{11}=6v^2U_1^2+8v^2U_1U_2+4v^2\Delta -12 U_1\Delta\, , 
 	\\
	&&\gamma_{1}^{22}=2\cos^2 \alpha \left[ v^2(2U_1+U_2)-4\Delta \right] \, ,
	\\
	&&\gamma_{1}^{12}=\gamma_{1}^{21}	= \cos^2\alpha\left[v^2(U_1+2U_2-\sin^2\alpha)-2\Delta\right] \, ,
	\\
	&&\gamma_{2}^{11}=2\sin^2\alpha\left[v^2(U_1+2U_2)-4\Delta\right] \, ,
	\\
	&&\gamma_2^{22}= 	6v^2U_2^2+8v^2U_1U_2+4v^2\Delta -12 U_2\Delta \, ,
	\\ 
 	&&\gamma_2^{12}=\gamma_2^{21}	=\sin^2\alpha	\left[v^2(2U_1+U_2-\cos^2\alpha)-2\Delta\right] 
\end{eqnarray*}
are the effective nonlinearities,
\begin{eqnarray*}
\begin{array}{l}
	\displaystyle{
	\beta_1^1= 
		v\Delta-\frac{v^3}{2}(U_1+2U_2),\quad \beta_1^2=\frac{v^3}{2}\cos^2\alpha\, ,
	}
	\\
	\displaystyle{
	 \beta_2^2=v\Delta-\frac{v^3}{2}(2U_1+U_2),\quad \beta_2^1=\frac{v^3}{2}\sin^2\alpha  
	 }
\end{array}
\end{eqnarray*}
are the effective dispersions, and the Einstein summation rule over repeated indexes is used.

Let us return to discussion of the integrable limit (\ref{manakov}), where according to the discussion following (\ref{amp-phase}), one must take $v=v_+=\sqrt{2(U_1+U_2)}$ (i.e. the upper branch only) and $\Delta=0$. In this case, the coefficients of the coupled KdV equations become $\gamma_{i}^{jk}=(U_1+U_2)\tilde{\gamma}_{i}^{jk}$ and $\displaystyle{\beta_{i}^{j}=\frac{v}{2}(U_1+U_2)\tilde{\beta}_{i}^{j}}$ where 
\begin{eqnarray*}
	\label{exact}
	&&\tilde{\beta}_1^1=-2(U_1+2U_2), \quad \tilde{\beta}_1^2=1+\sqrt{1-4U_1U_2}\, ,
	\\
	&&\tilde{\beta}_2^2=-2(2U_1+U_2), \quad  	\tilde{\beta}_2^1=1-\sqrt{1-4U_1U_2}\, ,
	\\
	&&\tilde{\gamma}_1^{11}= \frac{1}{9}(2\tilde{\beta}_2^2-\tilde{\beta}_1^1 )(5\tilde{\beta}_1^1+2\tilde{\beta}_2^2)\, ,
	\\
	&&\tilde{\gamma}_1^{12}=\tilde{\gamma}_1^{21}=	-\frac 12 \tilde{\beta}_1^2(\tilde{\beta}_1^1+\tilde{\beta}_2^1)\, ,
	\\
	&&\tilde{\gamma}_1^{22}=- \tilde{\beta}_2^2\tilde{\beta}_1^2, 	\quad	\tilde{\gamma}_2^{11}=- \tilde{\beta}_1^1\tilde{\beta}_2^1\, ,
	\\
	&&\tilde{\gamma}_2^{12}=	 \tilde{\gamma}_2^{21}= 	 -\frac 12 \tilde{\beta}_2^1(\tilde{\beta}_2^2+\tilde{\beta}_1^2)\, ,
	 \\
	&&\tilde{\gamma}_2^{22}= \frac{1}{9}(2\tilde{\beta}_1^1-\tilde{\beta}_2^2 )(5\tilde{\beta}_2^2+2\tilde{\beta}_1^1)
\end{eqnarray*}
(notice that now $U_1U_2\leq 1/4$).
The respective system of the coupled KdV equations is integrable.

\subsection{Small amplitude dark solutions}

A specific particular solution of Eq.~(\ref{kdv}) can be searched in the form
\begin{eqnarray}
\label{aa}
	q_1= q_2=-\frac{\eta}{\cosh^2\left[\sqrt{\frac2v}\kappa(\zeta-w\tau)\right]}
\end{eqnarray}
where $w$, and $\eta$ are real constants to be determined and $\kappa$ is a constant parameterizing the problem.  Substitution of ansatz (\ref{aa}) in Eq.~(\ref{kdv}) gives the condition of the equality of the chemical potentials 
\begin{eqnarray}
\label{connect}
	{\cal E}_1={\cal E}_2\,.
\end{eqnarray}
Next, the parameter $\eta$, characterizing the width of the soliton, is computed to be
\begin{eqnarray}
\label{eta}
	\eta=\frac{2\kappa^2}{1+U_1+U_2}.
\end{eqnarray}
The parameter $w$, which characterizes the soliton velocity in the frame moving with the speed of the sound, is given by
\begin{eqnarray}
	\label{w+}
	w=w_+=-4(U_1+U_2)(1+U_1+U_2)\kappa^2
\end{eqnarray}
for the upper branch and 
\begin{eqnarray}
	\label{w-}
	w=w_-=-4(U_1+U_2-1)^2\kappa^2
\end{eqnarray}
for the lower branch of the spectrum.
Both $w_\pm$ are negative, what means that the solitons move with velocities smaller than the sound velocities of the respective branches.

As it follows directly from ansatz (\ref{aa}), solutions describing different branches correspond to equal distributions of atomic densities. 
Meantime they correspond to different phase differences 
\begin{eqnarray}
\label{phase_diff}
\Delta\varphi_j=\phi_{j+}-\phi_{j-} 
\end{eqnarray}
at the infinities [see Eqs. (\ref{amp-phase}) and (\ref{boundary1})]: the lower-branch soliton ``separates'' domains with larger difference of chemical potentials.

\subsection{Numerical results}

As it has been mentioned above, although solutions of the coupled KdV equations represent a good approximation for the exact solutions of the coupled NLS equation in the small amplitude limit, strictly speaking they cannot be considered as satisfactory, when applied to the dynamics of a two-component BEC in an elongated trap. Meantime, the obtained vector soliton (\ref{aa}) can be employed as an initial condition for numerical generation of dark vector solitons of the coupled NLS equations (\ref{NLS-dimless}),  having small but finite amplitudes. Such numerical study is performed in the present section. 

Taking into account the existence of two kinds of small amplitude solitons we address  the question about persistence of the excitations at finite, but still small, amplitudes, as well as their stability withing the framework of system (\ref{NLS-dimless}).

An exact  vector dark soliton corresponding to the upper branch of the linear spectrum reads
\begin{eqnarray}
	\label{dark1}
	\psi_j=\left(iv_0+\sqrt{1-v_0^2}\tanh\left[\kappa(x-V_0t)\right]
	\right)e^{-i{\cal E}_jt}
\end{eqnarray}
where  
\begin{eqnarray}
	V_0=\sqrt{1+U_1+U_2-4\kappa^2},\quad v_0=\frac{V_0}{\sqrt{1+U_1+U_2}}
\end{eqnarray}
and relation  (\ref{connect}), which also can be rewritten as  $2\cos^2\alpha=1-U_1+U_2$, 
is taken into account. 
By expanding (\ref{dark1}) in the Taylor series in terms of the small parameter $\kappa$ one verifies that the leading orders transform into the upper-branch dark soliton described by the formulas (\ref{expan1}), (\ref{expan2}),  (\ref{aa}), (\ref{eta}), and (\ref{w+}), where the formal small parameter $\varepsilon$ is substituted by one and $\kappa$ is interpreted as the small parameter of the problem. 
	
\begin{figure}[t]
\includegraphics[width=5cm,clip]{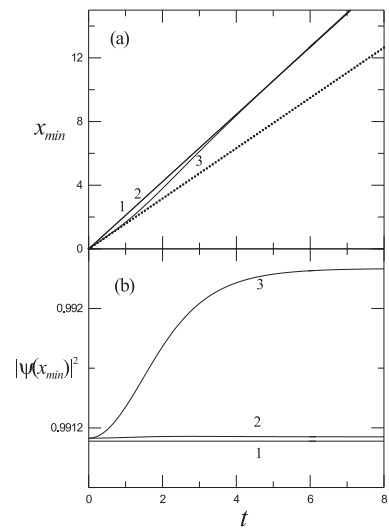}
\caption{(a) Trajectories of dark solitons. Lines 1, 2 and 3 correspond to dynamics of  $x_{min}^{(0)}$, $x_{min}^{(+)}$ and $x_{min}^{(-)}$ 
computed numerically. The dotted line corresponds to the theoretical prediction of dynamics of the $x_{min}^{(-)}$. 
Theoretically predicted values of $x_{min}^{(0)}$ and $x_{min}^{(+)}$ practically coincide, what makes them indistinguishable in the figure. Parameters are $U_1=2$, $U_2=1.5$, $\alpha=\pi/3$, $\kappa=0.1$, $\varepsilon=1$.
(b) Dynamics of the minima of densities of the dark vector solitons: the grows of the minimum shown by the line 3 corresponds to shallowing the soliton solution.}
\label{fig1}
\end{figure}

In Fig.\ref{fig1}(a) the trajectories of the centers of vector dark solitons are shown for three different initial conditions: the exact dark soliton (\ref{dark1}) (line 1), the approximate distribution (\ref{aa}) corresponding to the upper (line 2), and lower (line 3) branches. The centers of the solutions are defined as coordinates of the absolute minima of $|\psi_j|^2$: they are designated as $x_{min}^{(0)}$ and $x_{min}^{(\pm)}$ for the exact and two approximate solutions, respectively. 

\begin{figure}[t]
\includegraphics[width=5cm,clip]{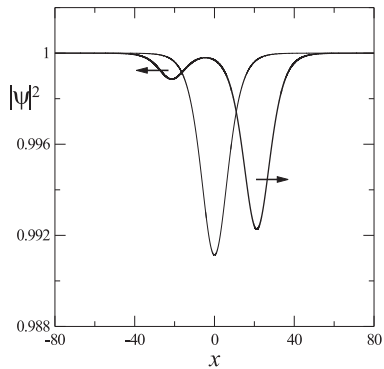}
\caption{Profile of the unstable lower-branch soliton  at $t=0$ (thin line) and at $t=10$ (thick line). Parameters are the same as in Fig.\ref{fig1}.
}
\label{fig2}
\end{figure}

Exact solution (\ref{dark1}) appears to be stable, and its approximate counterpart given by (\ref{aa}) undergoes small (invisible on the scale of Fig.\ref{fig1}) deformation. In the numerical simulation with the ``approximate'' initial condition corresponding to (\ref{aa}) the effective small parameter $\kappa$ is $0.1$. This gives the amplitude difference with the exact solution to be of order of $\kappa^4=10^{-4}$ what explains difference between the lines 1 and 2 in Fig.\ref{fig1}(b). The solution corresponding to the lower branch of the spectrum starts to move with the velocity $v_-$, but during initial interval of time changes significantly: it decays into two localized pulses moving in opposite directions, as it is shown in Fig.\ref{fig2}. The forward moving part represents the ``unstable'' dark soliton. Its amplitude continues to change during the evolution (see Fig.\ref{fig1}(b)) and its velocity approaches the group velocity of the upper (stable) branch (see Fig.\ref{fig1}(a)). 

The observed instability of the lower branch dark soliton is especially interesting in view of the recent results reported in \cite{Berloff}, where it was found that in the 3D coupled GP equations, describing spinor condensates, there exist subsonic (i.e. having velocities below the sound speed of the lowest branch) solitary wave complexes, most of which are suggested to be stable. In our case a subsonic vector dark soliton is unstable, and the stable one has the velocity between the two sound speeds: $v_-<V_0<v_+$.

In order to understand this phenomenon, let us consider the energy of the system  
\begin{eqnarray}
	\label{energy}
	E=E_1+E_2+E_{int}
\end{eqnarray}
which we write down in dimensionless variables
\begin{eqnarray}
	\label{energy1}
	E_j= \int dx\left[|\psi_{jx}|^2+\frac {U_j}{2}\left(|\psi_j|^2-1\right)^2\right]
\end{eqnarray}
is the energy of the excitation of $j$-th component and
\begin{eqnarray}
\label{energy2}
	E_{int}=\frac 12\int dx\left(|\psi_1|^2-1\right)\left(|\psi_2|^2-1\right)
\end{eqnarray}
is the energy of interaction of the components.
Next, we recall that for a given phase differences between the infinities, $\Delta\varphi_j$ [see definition (\ref{phase_diff})], there exist two types of dark solitons, corresponding to two branches of the group velocities. Let us assume now that the respective lower and upper branch solitons are characterized by parameters $\kappa_-$ and $\kappa_+$, respectively.  Let also initially the low branch soliton is excited. Since it does not represent an exact (but only an approximate) solution it starts to deform with time. Such a deformation has a constrain: the phase differences $\Delta\varphi_j$ are preserved. In the case of a small amplitude soliton, given by (\ref{aa}), one computes from Eq.~(\ref{amp-phase}) and definition (\ref{phase_diff})
\begin{eqnarray}
\label{eqv_phase}
	\Delta\varphi_1&=&\Delta\varphi_2=\nonumber\\ 
\frac{2^{3/2}}{1+U_1+U_2}\kappa_+v_+^{3/2}&=&\frac{2^{3/2}}{1+U_1+U_2}\kappa_-v_-^{3/2}.
\end{eqnarray}
 
On the other hand, one can compute the energy in terms of the parameters of the solution
  by  substitution of (\ref{aa}) and (\ref{eta}) into expressions   (\ref{energy})-(\ref{energy2}):  
\begin{eqnarray}
\label{energy_}
E_\pm=\frac{16\kappa_\pm^3  \left[v_\pm^2+2(1+U_1+U_2)\right]}{3(1+U_1+U_2)^2}.
\end{eqnarray}
From (\ref{eqv_phase}) it follows that $\kappa_+v_+^{3/2}=\kappa_-v_-^{3/2}$. Since  $v_-<v_+$ one finds that $\kappa_+<\kappa_-$ and hence $E_+<E_-$.
Thus what we observed in the numerical simulations is the transformation of a high energy vector soliton in a low energy vector soliton, accompanied by quasilinear modes.

\section{Effect of the  parabolic trap}

Let us now turn to the situation when the trap is not long enough in the axial direction, such that reflections of a soliton from potential walls can happen. In this case the trap potential must be included explicitly in the equations (instead of including it into the normalized ground states, see e.g.~\cite{reviewBK}). This leads to coupled 1D GP equations (\ref{NLS-dimless}) as follows
\begin{eqnarray}
	\label{NLS-dimless1}
	\begin{array}{l}
	i\partial_t\psi_1=-\partial_{x}^2\psi_{1}+(\nu_1^2 x^2+U_1|\psi_1|^2+\cos^2\alpha|\psi_2|^2)\psi_1\,,
	\\
	i\partial_t   
	\psi_{2}=-\partial_{x}^2\psi_{2}+(\nu_2^2 x^2+ \sin^2\alpha|\psi_1|^2+U_2|\psi_2|^2)\psi_2\,.
	\end{array}
\end{eqnarray}
Here $\nu_1=\lambda /\chi\rho^2$ and $\nu_2=\omega\nu_1$ are effective strengths of parabolic traps. Without restriction of generality in what follows we consider the situation where $\nu_1<\nu_2$.

In the case of equal effective frequencies, $\nu_1=\nu_2$, and subject to the specific initial condition ($\psi_1=$const$\cdot\psi_2$), effectively reducing the system to a single NLS equation, a vector dark soliton oscillates with the frequency $\sqrt 2$ times smaller than the frequency of the parabolic trap~\cite{BA1,BK03}. If however
strengths of the parabolic potentials of the two components are different,  behavior of the soliton changes dramatically. There exist two competing factors, which determine the dynamics. On the one hand, each component, affected by its own parabolic potential, ``attempts'' to oscillate with its own frequency. On the other hand, attractive interaction between the components, described by functional (\ref{energy2}), forces them to oscillate with the same frequency. 

\begin{figure}[ht]
\includegraphics[width=7cm,clip]{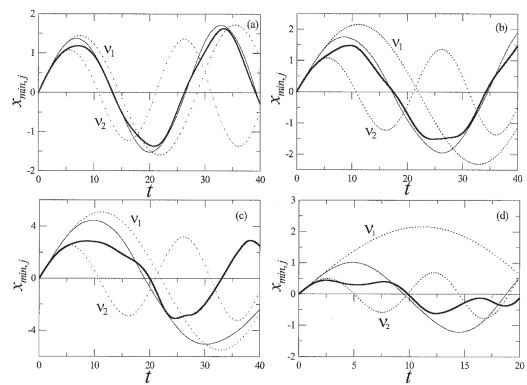}
\caption{Trajectories of the components of the vector soliton, $x_{min,j}$ ($j=1,2$), corresponding to the first (thin line) and to the second (thick line) components for  parameters $\kappa$, $\nu_1$ and $\nu_2$ being respectively:  (a) $1.05$, $0.15$, and $0.2$; (b) $1.05$, $0.1$, and $0.2$; and (c) $1.$, $0.1$, and $0.2$; 
(d) $1.05$, $0.2$, and $0.4$. 
Dotted lines show trajectories of the components in the respective traps when interaction between the components is absent. The other parameters are $U_1=2$, $U_2=1.5$, and $\alpha=\pi/3$.
}
\label{fig3}
\end{figure}

The dynamics emerging from the competition of the factors mentioned above is shown in  Fig.\ref{fig3}.  In the case of sufficiently wide potentials and small difference between their strengths (Fig.\ref{fig3}(a)) the first and second components (thin and thick lines correspondingly) initially oscillate with approximately equal frequencies. 

The dynamics observed in the case when initial soliton velocities are fixed  and  difference between strengths of the traps  is increased is shown in Fig.\ref{fig3}(b). One observes more significant separation of the components. The dynamics, however still resembles periodic motion. After subsequent increase of the initial kinetic energy of the soliton, by means of increasing its initial velocity,  more visible splitting is observed, as it is shown in Fig.\ref{fig3}(c). After the first period the soliton splits and each component starts to oscillate with its own frequency.

By increasing the frequencies of the magnetic trap but keeping constant the ratio between them (this corresponds to passage from  Fig.\ref{fig3}(b) to Fig.\ref{fig3}(d)) one observes fast splitting the components whose trajectories show rather independent behavior. The second component, which exists in effectively more narrow trap, does not display periodic motion. 
 
To understand qualitatively the described behavior, we introduce coordinates of the centers of mass of the components ($j=1,2$)
\begin{eqnarray}
X_j=\frac{1}{N_j}\int\limits_{-\infty}^{\infty}x |\psi_j(x)|^2dx,\quad N_j=\int\limits_{-\infty}^{\infty} |\psi_j(x)|^2dx\,.
\end{eqnarray}
We also define the coordinate of the center of mass of the whole condensate $X_+=(N_1X_1+N_2X_2)/N$, where $N=N_1+N_2$, and the distance between the two centers of masses: $X_-=X_2-X_1$. We emphasize that, strictly speaking, $X_{1,2}$ do not describe trajectories of the dark solitons (see the discussion in \cite{BK03}). One however could expect, that when the vector soliton splitting is small enough (i.e. when $X_-\ll 1$) the relations among the frequencies of the two-component problem are approximately the same as in the one-component case. That is why we concentrate on dynamics of $X_\pm$. 

Differentiating $X_j$ with respect to time and using (\ref{NLS-dimless}) we compute
\begin{eqnarray}
	\label{X+}
	\ddot{X}_+&+&\frac 4N \left(N_1\nu_1^2+N_2\nu_2^2\right)X_+\nonumber \\
	&+&\frac 4N \left[N_2\left(1-\frac{N_2}{N}\right)\nu_2^2-\frac{N_1N_2}{N}\nu_1^2\right]X_-
\nonumber \\	&=&\frac 2N\left(\sin^2\alpha-\cos^2\alpha\right)\int\limits_{-\infty}^{\infty}|\psi_1|^2\frac{\partial |\psi_2|^2}{\partial x}dx,
\\
\label{X-}
	\ddot{X}_-&+&4\left[\left(1-\frac{N_2}{N}\right)
	\nu_2^2+\frac{N_2}{N}\nu_2^2\right]X_-+4 \left(\nu_2^2-\nu_1^2\right)X_+
\nonumber \\	&=&2\left(\frac{\sin^2\alpha}{N_2}+\frac{\cos^2\alpha}{N_1}\right)\int\limits_{-\infty}^{\infty}|\psi_1|^2\frac{\partial |\psi_2|^2}{\partial x}dx\,.
\end{eqnarray}
Hereafter overdots  stand for derivatives with respect to time.

While equations (\ref{X+}) and (\ref{X-}) are exact, they are not closed, and in order to make use of them we have to make some approximation. Below, in Figs.~\ref{fig5}(a),(b) we show that, if the difference $\nu_2-\nu_1$ is small enough, the shapes of the components are preserved for relatively long temporal intervals, even when the splitting is not negligible. Basing on this observation we assume that the components of the vector soliton preserve their shapes and only change their velocities. This allows to us describe each component in the vicinity of the soliton by formula (\ref{dark1}) where $V_0t$ is substituted by $X_j(t)$. We also restrict the consideration (this time for the sake of simplicity only) to zero initial velocities: $v_0=0$. Then the integral in the  right hand sides of (\ref{X+}) and (\ref{X-}) can be computed explicitly. It appears to be a function of $X_-$ only (i.e. independent on $X_+$). 

\begin{figure}[ht]
\includegraphics[width=8cm,clip]{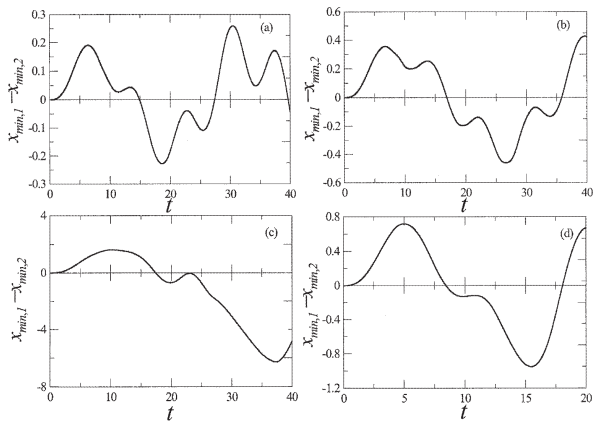}
\caption{Dynamics of $x_{min,1}-x_{min,2}$ corresponding to Fig.\ref{fig3}. The panels (a)--(d) correspond to the respective panels in Fig.\ref{fig3}.
}
\label{fig4}
\end{figure}

\begin{figure}[ht]
\includegraphics[width=8cm,clip]{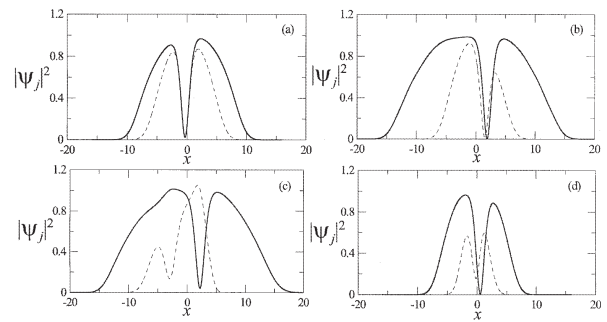}
\caption{Density profiles $|\psi_j|^2$ of the first (solid lines) and second (dashed lines) components at time $t_f$ corresponding to Fig.\ref{fig3}. In (a)--(c) $t_f=40$ and in (d) $t_f$=20.
}
\label{fig5}
\end{figure}

Let us now take into account that $|X_-|\ll |X_+|$ [see Figs.~\ref{fig3} and~\ref{fig4}, panels (a) and (b)]. Then, in the leading order the terms with $X_-$ can be neglected in (\ref{X+}) resulting in a simple equation for the center of mass of the condensate:
\begin{eqnarray}
\label{X++}
	\ddot{X}_++\Omega_+^2X_+=0. 
\end{eqnarray}
Here
\begin{eqnarray}
\label{main_friq}	
	\Omega_+^2=\frac 4N \left(N_1\nu_1^2+N_2\nu_2^2\right).
\end{eqnarray}
In other words $\Omega_s=\Omega_+/\sqrt{2}$ is expected to be the main frequency of the oscillations of the vector soliton, where we introduce the factor $\sqrt{2}$ conjecturing that the relation between the frequencies of the condensate and the soliton is the same as in the one-component case. To compare this estimate with the direct numerical simulations shown in Fig.\ref{fig3}(a), we use the numerical values $N_1\approx 12$ and $N_2\approx 7.1$ and obtain from (\ref{main_friq}):  the frequency $\Omega_s\approx 0.24$ and the period of oscillations $T_s\approx 26.1$. The numerical value of the period subtracted from Fig.\ref{fig3}(a) is $T_{num,s}\approx 26.3$ what is in a remarkable (taking into account the character of the approximations) agreement with the theoretical prediction $T_s$.

In order to analyse the characteristics of the splitting of the soliton, we hold the above assumptions and rewrite Eq.~(\ref{X-}) in the form
\begin{eqnarray}
\label{X--}
	\ddot{X}_-+\Omega_-^2X_-+\frac{\partial U_{eff}(X_-)}{\partial X_- }=
	4 \left(\nu_1^2-\nu_2^2 \right)X_+
\end{eqnarray}
where the effective potential is given by
\begin{eqnarray}
U_{eff}(X_-)&=& 8\left(\frac{\sin^2\alpha}{N_2}+\frac{\cos^2\alpha}{N_1}\right)
\nonumber \\
&\times& \frac{\sinh X_--X_-\cosh X_-}{\sinh^3X_-} \,.
 \end{eqnarray}
$U_{eff}(X_-)$ describes additional confinement of the relative motion of the components due to the attractive interaction between the components, which is imposed  simultaneously with the parabolic trap characterized by the frequency
\begin{eqnarray}
	\label{omega_second}
	\Omega_-^2=4\left[\left(1-\frac{N_2}{N}\right)
	 \nu_2^2+\frac{N_2}{N}\nu_2^2\right].
\end{eqnarray}
Eq.~(\ref{X--}) is nothing but a nonlinear oscillator in a trap made up of the  parabolic potential $\Omega_-^2X_-^2/2$ and of the nonlinear potential $U_{eff}(X_-)$, driven by the periodic force, $4 \left(\nu_1^2-\nu_2^2 \right)X_+(t)$, originated by the oscillation of the condensate as a whole.

To simplify the next consideration we take into account that $X_-$ is relatively small [see Fig.\ref{fig4}(a)] and substitute $\partial U_{eff}(X_-)/\partial X_- $ by the first term of its Taylor expansion. This results in a modified linear frequency, which is determined  as
\begin{eqnarray}
	\tilde{\Omega}_-=\sqrt{\Omega_-^2+\frac{\partial^2 U_{eff}}{\partial X_-^2 }(0)}.
\end{eqnarray}

Using the data of Fig.\ref{fig3}(a) we compute $\tilde{\Omega}_-\approx 0.59$, what corresponds to the period of modulation $T_-=10.65$, while the numerical value subtracted from the figure is $T_{num,-}\approx 11.8$. Thus we again observe good agreement between the simple theoretical estimates and the numerical results, what shows that the simple model given by Eqs.~(\ref{X++}) and (\ref{X--}) provides adequate description of the vector soliton dynamics in a parabolic trap whenever the splitting between the components is small. 
  
Interaction between the components of the vector soliton and soliton interaction with the confining potential in a general case lead to deformations of the profiles of the soliton components, what is shown in   Fig.\ref{fig5} for the times $t_f$ corresponding to the final time of the evolution depicted in Fig.\ref{fig3}. Already after few oscillations the vector soliton decays, 
except in the case when both components oscillate in effective traps with equal strengths (i.e. when the vector soliton behaves like a single-component dark soliton \cite{BK03}). 

\section{Conclusion}

To conclude, we have investigated dynamics of vector dark solitons governed by one-dimensional coupled nonlinear Schr\"{o}dinger equations. 
In the homogeneous case and in the small amplitude limit, when the vector soliton propagates with the velocity close to the speeds of the sound, a stable vector soliton has a velocity close to the higher velocity of the sound and exceeding speed of the slow sound. The respective subsonic (i.e. having a velocity lower than the speed of the slow sound) dark vector soliton is unstable. The both cases are described by the coupled Korteweg-de Vries equations. When the group velocity of the lower branch of the sound dispersion relation becomes zero, what corresponds to the integrable Manakov system, the coupled Korteweg-de Vries equations obtained in the small amplitude limit are integrable (to the best of authors knowledge, such system has not been reported in the literature, so far). 

Including a parabolic trap into consideration changes behavior of vector solitons dramatically, leading in a general case to their decay, what is explained by different eigenfrequencies of the two components. If meantime the effective traps for the both components have close frequencies, during initial times the dynamics of the vector soliton can be qualitatively (and also quantitatively, with rather good accuracy) described by the oscillatory motion of the soliton center with the frequency given by (\ref{main_friq}).
Relative dynamics of the components, when splitting is small enough, can be interpreted as an oscillator driven by a periodic force.  
Large difference between strengths of parabolic traps or large initial soliton velocities cause instability of dark vector solitons leading to their splitting and subsequent decay, thus preventing possibility of the long-time dynamics.

\acknowledgments

Authors thank L. P. Pitaevskii, V. S. Gerdjikov and V. E. Vekslerchick for useful comments. V.A.B. acknowledges support by the FCT fellowship SFRH/BPD/5632/2001.
The work was partially supported by the grant POCTI/FIS/56237/2004.

\bigskip

\appendix

\section{Derivation of the coupled KdV equations}

Substituting (\ref{sol_expan}), (\ref{expan1}), (\ref{expan2}) and (\ref{CP}) in (\ref{NLS-dimless}), and gathering terms of the same order of the small parameter $\varepsilon$ one obtains that the equation is satisfied identically in the orders $\varepsilon^0$ and $\varepsilon^1$. Next one has
\begin{eqnarray}
	\label{app:1}
	\begin{array}{l}
	v\partial_\zeta\phi_1-2U_1q_1-2\cos^2\alpha \,q_2=0
		\\
		v\partial_\zeta\phi_2-2\sin^2\alpha\, q_1- 2U_2q_2 =0
	\end{array}
\end{eqnarray}
in the order $\varepsilon^2$ and
\begin{eqnarray}
	\label{app:2}
	\partial_\zeta^2\phi_1-v\partial_\zeta q_1=0,\quad \partial_\zeta^2\phi_2-v\partial_\zeta q_2=0 
\end{eqnarray}
in the order $\varepsilon^3$. Integrating equations (\ref{app:2}) once and taking into account (\ref{boundary1}), one obtains Eq.~(\ref{amp-phase}). Substituting link (\ref{amp-phase}) in (\ref{app:1}) one verifies that the so obtained system is solvable subject to the condition (\ref{group}), what justifies the value of the soliton velocity as the group velocity of the sound waves.

Equations of the order $\varepsilon^4$, where the link (\ref{amp-phase}) is used, read
\begin{widetext}
\begin{eqnarray}
	\label{app:4}
	\begin{array}{l}
\displaystyle{
	v\partial_\zeta\phi_{11}-2U_1q_{11}-2\cos^2\alpha\, q_{21}=\partial_\tau\phi_1- 
	\frac 1v \partial_\zeta^3\phi_1
	+\frac{3U_1}{v^2}(\partial_\zeta\phi_1)^2+
	\frac{2\cos^2\alpha}{v^2}(\partial_\zeta\phi_1)(\partial_\zeta\phi_2)
	+\frac{\cos^2\alpha}{v^2}(\partial_\zeta\phi_2)^2
	}	\, ,
		\\
	\displaystyle{	
	v\partial_\zeta\phi_{21}-2\sin^2\alpha\, q_{11}- 2U_2q_{21}=\partial_\tau\phi_2- 
	\frac 1v \partial_\zeta^3\phi_2
	+\frac{3U_2}{v^2}(\partial_\zeta\phi_2)^2+
	\frac{2\sin^2\alpha}{v^2}(\partial_\zeta\phi_1)(\partial_\zeta\phi_2)
	+\frac{\sin^2\alpha}{v^2}(\partial_\zeta\phi_1)^2
	}\, .
	\end{array}
\end{eqnarray}
\end{widetext}
Finally one computes the equations of the order $\varepsilon^5$ (where the link (\ref{amp-phase}) as well as the explicit value of the group velocity (\ref{group}) are taken into account and integration with respect to $\zeta$ is performed)
\begin{eqnarray}
	\label{app:5}
	\begin{array}{l}
\displaystyle{
\partial_\zeta\phi_{11}-v q_{11}=-\frac{3}{2v}(\partial_\zeta\phi_1)^2-\frac 1v \partial_\tau\phi_1
}\, ,
\\
\displaystyle{
\partial_\zeta\phi_{21}-v q_{21}=-\frac{3}{2v}(\partial_\zeta\phi_2)^2-\frac 1v \partial_\tau\phi_2
}\, .
\end{array}
\end{eqnarray}

The last system allows one to expresses $q_{j1}$ through other dependent variables, and  substitute the result in equations (\ref{app:4}). In this way one obtains the inhomogeneous linear algebraic system of equations for $\partial_\zeta\phi_{j1}$, which determinant is computed to be zero.  This leads to linear dependence of the equations which are solvable only subject to the respective requirement imposed on their right hand sides. Differentiating the mentioned solvability condition with respect to $\zeta$ and using one more time the link (\ref{amp-phase}) in order to express the final result through the dependent variables $q_j$ only, one obtains coupled KdV equations (\ref{kdv}).

\end{document}